# Anisotropy of the upper critical field in a Co-doped BaFe$_2$As$_2$ single crystal

M. Kano[1], Y. Kohama[2,3], D. Graf[1], F. F. Balakirev[2], A. S. Sefat[4], M. A. McGuire[4], B. C. Sales[4], D. Mandrus[4] and S. W. Tozer[1]

[1] *National High Magnetic Field Laboratory, Tallahassee, FL 83810, USA*
[2] *MPA-NHMFL, Los Alamos National Laboratory, Los Alamos, New Mexico 87545, USA*
[3] *Materials and Structures Laboratory, Tokyo Institute of Technology, 4259 Nagatsuta-cho, Midori-ku, Yokohama 226-8503, Japan*
[4] *Materials Science and Technology Division, Oak Ridge National Laboratory, Oak Ridge, TN 37931, USA*

Abstract

The temperature dependence of the upper critical magnetic field ($H_{c2}$) in a BaFe$_{1.84}$Co$_{0.16}$As$_2$ single crystal was determined via resistivity, for the inter-plane ($H\perp ab$) and in-plane ($H//ab$) directions in pulsed and static magnetic fields of up to 60 T. Suppressing superconductivity in a pulsed magnetic field at $^3$He temperatures permits us to construct an $H$-$T$ phase diagram from quantitative $H_{c2}(0)$ values and determine its behavior in low temperatures. $H_{c2}(0)$ with $H//ab$ ($H_{c2//}(0)$) and $H\perp ab$ ($H_{c2\perp}(0)$) are ~ 55 T and 50 T respectively. These values are ~ 1.2 - 1.4 times larger than the weak-coupling Pauli paramagnetic limit ($H_p$ = 1.84 $T_c$), indicating that enhanced paramagnetic limiting is essential and this superconductor is unconventional. While $H_{c2//}ab$ is saturated at low temperature, $H_{c2}$ with $H\perp$ab ($H_{c2\perp}$) exhibits almost linear temperature dependence towards $T$ = 0 K which results in reduced anisotropy of $H_{c2}$ in low temperature. The anisotropy of $H_{c2}$ was ~ 3.4 near $T_c$, and decreases rapidly with lower temperatures reaching ~ 1.1 at $T$ = 0.7 K.

Introduction

The discovery of high transition temperature ($T_c$) superconductivity in iron arsenide based compounds has generated great interest among superconductivity researchers [1]. The ternary iron arsenide, BaFe$_2$As$_2$ with the ThCr$_2$Si$_2$-type (122-type) structure as a new parent compound [2] was proposed. In $A$Fe$_2$As$_2$ ($A$ = Ca, Sr, Ba), hole-doping has yielded a $T_c$ of as high as 38 K [3]. Partial substitution of the iron by cobalt ions in BaFe$_2$As$_2$ induced electron-doped superconductivity with an onset transition temperature $T_c^{onset}$ = 22 K [4]. The parent compound BaFe$_2$As$_2$ shows a first-order structural phase transition from tetragonal to orthorhombic with the simultaneous onset of long-range antiferromagnetic order around 140 K [5]. The superconducting state appears in the vicinity of these transitions either by applying pressure or introducing electron- or hole-doping in these series, suggesting that the antiferromagnetic spin fluctuation of Fe plays an important role in developing the superconducting ground state [2]. The paring mechanism in iron arsenide pnictides is consistent with the multi-band theory, and the so-called "extended s±-wave" model with a sign reversal of the order parameter between different sheets of Fermi surface [6].

Exploring the temperature dependence of $H_{c2}$ and its anisotropy, which can be estimated by transport measurements, are very important factors which help reveal the mechanism of superconductivity experimentally. High transition temperatures promise extremely high $H_{c2}$ values. High temperature $H_{c2}(T)$ curves are often extrapolated to obtain low temperature $H_{c2}(0)$ values. However, in other anisotropic superconductors such as organics [7], the low-field temperature dependence of $H_{c2}(T)$ is often a very poor guide to both its detailed behavior at low temperatures and its finite value at $T$ = 0 K. Therefore transport measurements at extremely high magnetic fields and low temperatures near to $T$ = 0 K are necessary to explore the overall shape of the $H$-$T$ phase diagram. In the reports on LaFeAsO$_{0.89}$F$_{0.11}$, the upturn in curvature for $H_{c2\perp}$ was observed, which provides evidence of two-band superconductivity [8] On (Ba, K)Fe$_2$As$_2$ which is hole doped, the nearly isotropic shape of $H_{c2}$ was presented and showed that orbital limiting might be dominant at all field angles [9], and on Sr(Fe, Co)$_2$As$_2$ which is electron doped as our present material showed similar anisotropy in $Hc_2$ [10]. The anisotropy of $H_{c2}$ in the whole range of compositions in Ba(Fe,Co)$_2$As$_2$ was studied earlier [11]. Then the anisotropy of $Hc_2$ with bulk specific-heat measurement was studied in BaFe$_{1.852}$Co$_{0.148}$As$_2$ [12], and in BaFe$_{1.8}$Co$_{0.2}$As$_2$, $Hc_2$ was presented up to 45 T, indicating small anisotropy and the low temperature extrapolation suggests $H_{c2//}$ ~ 70 T [13].

In this report, we present electrical resistivity measurements and its anisotropy in a BaFe$_{1.84}$Co$_{0.16}$As$_2$ single crystal for higher magnetic fields up to 60 T, displaying the temperature dependence of $H_{c2}$ at temperatures down to 0.6 K, which gives a good picture of the overall shape of the *H-T* phase diagram without the extrapolation.

Experimental

A large single crystal of Co-doped BaFe$_2$As$_2$ was grown out of FeAs flux, as described elsewhere [4]. The size of the crystal we used for these measurements was ~ 1 × 1 × 0.25 mm$^3$. Temperature dependent electrical resistivity of the *ab*-plane ($\rho_{ab}$) was measured with a standard four-probe method. The contact resistance was ~ 50 Ω. The current and frequency used throughout the pulsed field measurements were 0.1 ~ 1 mA and 50 ~ 100 kHz, respectively. The magnetic field dependence of the $\rho_{ab}$ was measured at fixed temperatures using a 60 T capacitor bank-driven pulsed magnet at the NHMFL at Los Alamos National Laboratory. There is hysteresis between the up and down traces due to heating caused by the pulsed magnetic field. Data shown here are from the down sweep of the field as the decay time of the pulsed field is much longer than its rise time (5.75 ms rise time and 20 ms exponential decay), which generates less heating. The $\rho_{ab}$ at several fixed temperatures for low magnetic fields range (H ≤ 16 T) DC measurements with AC drive mode, which reverses the current to remove thermo EMFs, was performed on a Quantum Design Physical Property Measurement System (PPMS), with slow cooling and field sweeping rate, 0.5 K/min and 50 Oe/sec respectively. Electrical contacts were made by using silver paint, and Au wires were used. The current used for the PPMS was 1 mA.

Result and Discussion

Shown in Fig. 1 is the temperature dependence of $\rho_{ab}$ in BaFe$_{1.84}$Co$_{0.16}$As$_2$ at zero magnetic field. This compound shows the metallic behavior and linear temperature dependence down to 23K. The midpoint transition temperature is $T_c^{mid}$ = 22 K at 0 T. The transition width at 0 T is $\Delta T_c = T_c (90\%) - T_c (10\%) = 0.54$ K. The $\Delta T_c$ value is much smaller than the one reported for Ba$_{0.6}$K$_{0.4}$Fe$_2$As$_2$ earlier ($\Delta T_c$ = 1.5 K), which suggests that the present BaFe$_{1.84}$Co$_{0.16}$As$_2$ sample is of higher quality.

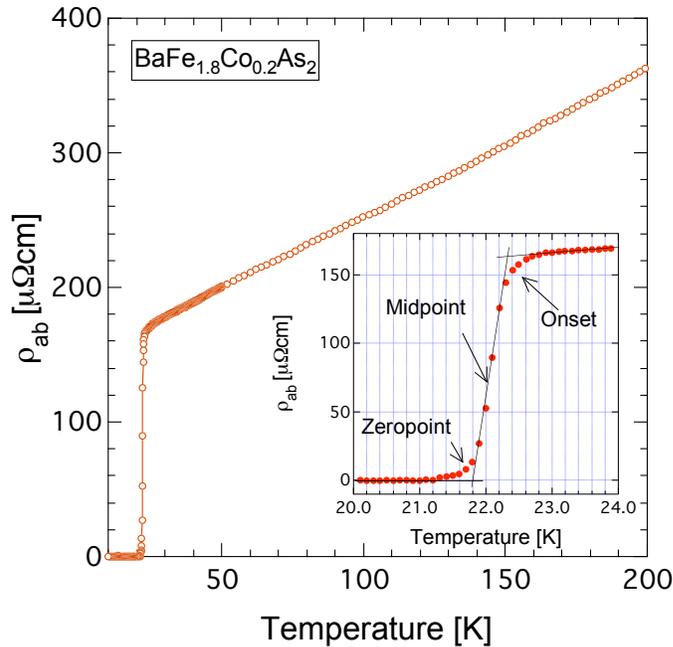

Fig.1. The temperature dependence of in-plane electrical resistivity ($\rho_{ab}$) of BaFe$_{1.84}$Co$_{0.16}$As$_2$ at zero magnetic field. The inset is a detailed view near the transition.

Resistivity data measured in pulsed magnetic fields up to 60 T at several fixed temperatures are shown in Fig. 2(a) and 2(b), for applied fields perpendicular to *ab*-plane ($H \perp ab$) and parallel to *ab*-plane ($H // ab$). The traces shown are normalized to the normal state value of each trace. The alignment along these orientations is accurate to within ± 3˚. As shown in Fig. 3, the resistance for increasing and decreasing fields from 0 to 16 T at several fixed temperatures was also measured in the PPMS and the data were consistent with that taken in the pulsed magnet. The superconducting transitions observed in this temperature range display no hysteresis.

In Fig. 2, both data show full recovery of the normal state resistivity above $H_{c2}$. The high field normal state of BaFe$_{1.84}$Co$_{0.16}$As$_2$ down to $^3$He temperatures was successfully observed, which is probably aided by the relatively low $T_c$. This allows us to determine quantitative values of $H_{c2}$ at $T = 0$ K and detailed behavior of the temperature dependence of $H_{c2}$ at low temperatures. The data show a sharp transition in $\rho_{ab}(H)$ as well as $\rho_{ab}(T)$ at zero field, which is an indication of high sample quality. A peak around 50 T in 0.7 K trace in Fig. 2 (b) is probably due to noise pick up.

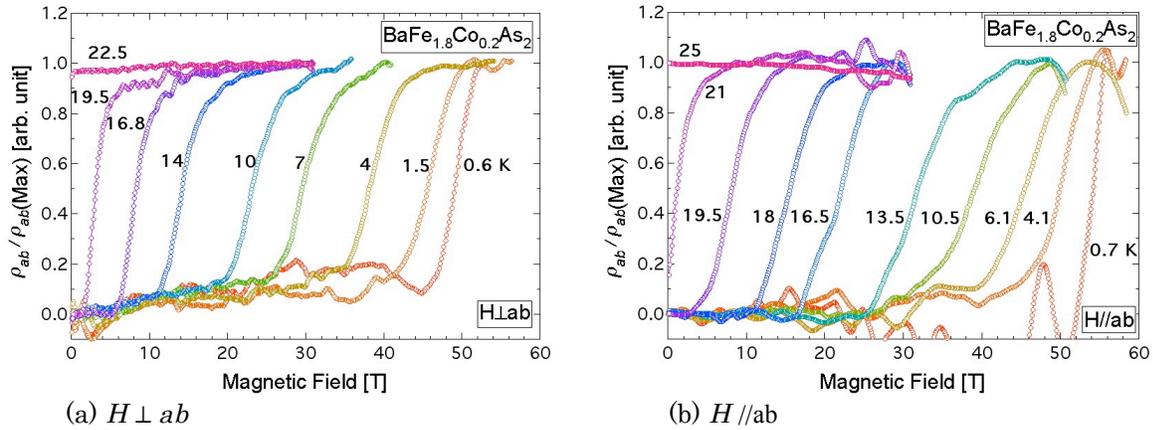

(a) $H \perp ab$  (b) $H //ab$

Fig. 2. In-plane resistance ($\rho_{ab}$) vs field for various temperatures in BaFe$_{1.84}$Co$_{0.16}$As$_2$, which was measured in pulsed magnetic fields. The applied magnetic fields were (a) perpendicular and (b) parallel to ab-plane, respectively. The temperature values in (a) are 0.6, 1.5, 4, 7, 10, 14, 16.8, 19.5, and 22.5 K, and in (b), those are 0.7, 4.1, 6.1, 10.5, 13.5, 15, 16.5, 18, 19.5, 21, and 25 K, respectively.

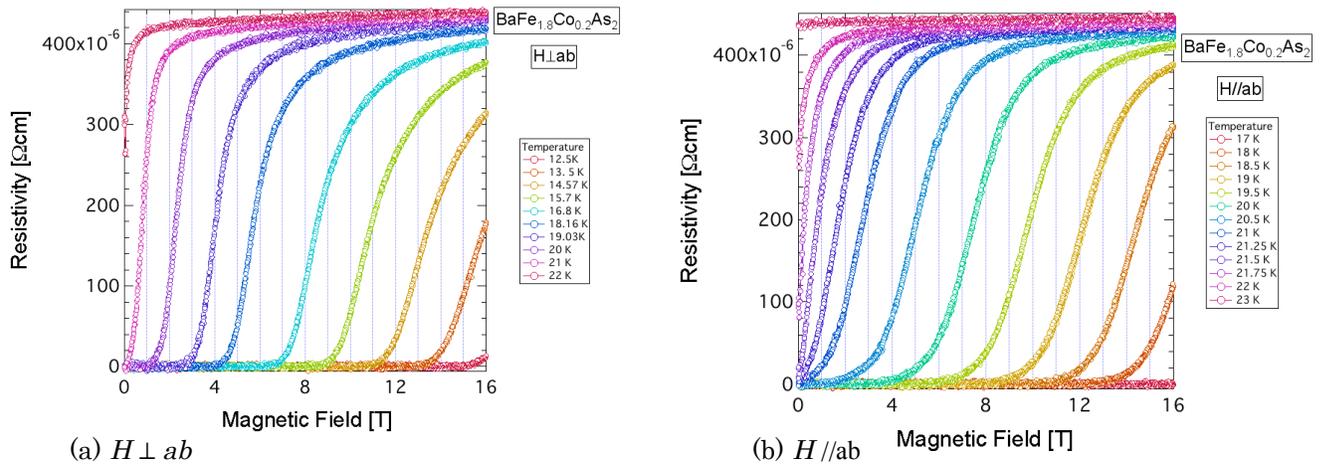

(a) $H \perp ab$  (b) $H //ab$

Fig. 3. In-plane resistance ($\rho_{ab}$) vs field for various temperatures in BaFe$_{1.84}$Co$_{0.16}$As$_2$, which was measured in the PPMS. The applied magnetic fields were (a) perpendicular and (b) parallel to ab-plane. The temperature values are as shown in the figures.

Fig. 4 shows the temperature dependence of the maximum in ($dR/dH$) measured in the PPMS, which demonstrates the sharpness of resistive transitions. Tinkham developed the resistive transition model based on flux dynamics [14], and has given an expression of the width and shape of the transition as follows: $R/R_n = (I_0[A(1-T/T_C)^{3/2}/2\mu_0 H])^{-2}$, where $I_0(x)$ is a modified Bessel function and $A$ is a constant proportional to the ratio of the zero field critical current to $T_c$. This model gives the temperature dependence of the transition width to be $\Delta H \propto (1-T/T_c)^{3/2}$. The decrease of the transition width is related to a strong pinning force. In our compound, the transition width decreases as the temperature increases for both orientations, which follows Tinkham's model. The superconducting transition in layered superconductors including high $T_c$ cuprates, $La_{1.85}Sr_{0.15}CuO_4$ [15], $YBa_2Cu_3O_x$ [16] and organic superconductors, $\kappa$-$(ET)_2Cu(NCS)_2$ [17], and $(TMTSF)_2PF_6$ [7] have similar temperature dependence as the present compound. $\rho_{ab}$ with the applied field of $H//ab$ is longitudinal, and $\rho_{ab}$ with the applied field of $H\perp ab$ is transverse magnetoresistance ($\rho_{ab}$), which should be more influenced by flux motion. Therefore the steep change in $dR/dT$ shown only for $H//ab$ is considered to be due to the force needed to push the flux according to the strong pinning force.

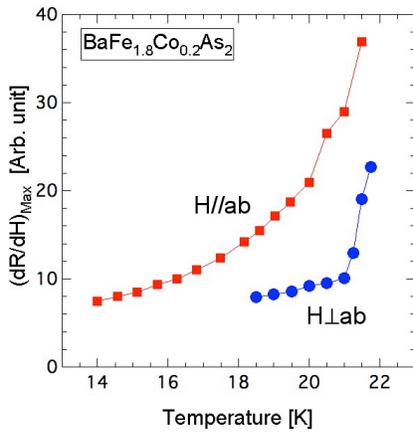

Fig. 4. The temperature dependence of the peak in derivatives of resistance curve ($dR/dH$) for both orientations; $H//ab$ and $H\perp ab$. Lines are guide to the eye.

Fig. 5 shows the temperature dependence of $H_{c2}$ in $BaFe_{1.84}Co_{0.16}As_2$, as determined from 50 % recovery of the resistance, with data taken from the down sweep of pulsed magnetic field and PPMS measurements, which are consistent with each other. In several highly two-dimensional superconductors including cuprates, the curvature of $H_{c2}(T)$ is reported to vary depending on the recovery percentage criteria used to determine $H_{c2}(T)$ [18]. In this compound, even if $H_{c2}$ is defined by 10 or 90 % recovery of the resistance, the shape of $H_{c2}$ curve does not change qualitatively. In the $H$-$T$ phase diagram at high temperatures near $T_c$, the quadratic behavior is observed for both field orientations. We believe this is due to flux dynamics as is seen in the cuprates [19], and is further supported with PPMS data.

$H_{c2}(T)$ clearly shows a difference in behavior depending on the direction of $H$. The curve of $H_{c2}(T)$ with $H//ab$ ($H_{c2//}$) has a tendency to saturate with decreasing temperature, while $H_{c2}(T)$ with $H\perp ab$ ($H_{c2\perp}$) presents a positive curvature near $T = 0$ K without saturation. The value of $H_{c2//}$ and $H_{c2\perp}$ at zero temperature are about 55 and 50 T, equivalent to 2.5 $T_c$ and 2.3 $T_c$, respectively. There are two independent mechanisms for suppressing superconductivity with magnetic fields; one is orbital pair breaking of cooper pairs in the superconducting state associated with screening currents generated to exclude the external field (orbital limit). The other one is a spin effect due to Zeeman splitting which applies only to the singlet pairings and limits superconductivity below the Pauli or Clongston-Chandrasekhar limit [20] as given by weak-coupling BCS paramagnetic formula, $H_p^{BCS}(T=0) = 1.84\, T_c(H=0)$ for isotropic s-wave pairing in the absence of spin-orbit scattering. In our case, we obtain $H_p^{BCS} = 40$ T and our $H_{c2//}$ is 1.4 times this limit, indicating that the Zeeman paramagnetic pair breaking may be essential for $H_{c2//}$. It is common that the weak-coupling BCS formula can underestimate the actual paramagnetic limit. Including strong electron - phonon coupling: $H_p \cong (1 + \lambda_{ep})H_p^{BCS}$ [Eq. 1][21], can significantly enhance $H_p$. If we take $\lambda_{ep} = 0.5$ which was obtained by simple estimation for $BaFe_2As_2$-based superconductors [22], Eq. 1 yields $H_p$ to be about 60 T, which is closer to our $H_{c2//}(0)$.

On the other hand, the $H_{c2\perp}$ shows positive curvature near $T = 0$ K without saturation. We consider that the residual upward curvature of $H_{c2\perp}(T)$ is originating from two-band features recently shown by various experiments in iron arsenide compounds [8]. The solid curve in Fig. 5 is a fit of the 2-band theoretical equation to $H_{c2\perp}$ [23]; $a_0[\ln t + U(h)][\ln t + U(hh)] + a_1[\ln t + U(h)] + a_2[\ln t + U(hh)] = 0$ [Eq. 2]. The $a_1$, $a_2$ and $a_3$ are determined from the BCS coupling constant tensor $l_{mm'}$, and the other parameters are defined by $U(x) = y(1/2 + x) - y(1/2)$, $h = H_{c2}D_1/2f_0T$, $t = T/T_c$ and $h = D_2/D_1$, where $y$ is the di-gamma function, $f_0$ is the magnetic flux quantum and $D_m$ are the electronic diffusivity for the $m^{th}$ Fermi Surface sheets. Here, we take the inter-band coupling values $l_{12} = l_{21} = 0.5$ from reference 7, and also use the same intra-band coupling value $l_{11} = l_{22} = 0.5$. We find that $H_{c2\perp}(0) = 49$ T and a small diffusivity ratio ($h$) of 0.085. The small $h$ implies stronger scattering in one of the bands. In fact, recent calculations show that the Co dopant induces stronger scattering in the hole band [4]. It should be noted that the exact determination of inter-band coupling values is impossible from our present data. When we use smaller $l_{11}$ ($l_{11} = 0.1$), Eq.2 also fits our $H_{c2}$ data very well with the smaller $h$ of 0.050 and $H_{c2\perp}(0)$ of 50 T. Therefore, we assume that the superconductivity in $BaFe_{1.84}Co_{0.16}As_2$ is explained well by the two-band model, in agreement with the fit to $H_{c2}$ curvature using the two-band theory.

Another possible explanation comes from a magnetic field-induced dimensional crossover (FIDC), as discussed by Lebed [24] and Dupuis, Montambaux, and Sá de Melo (DMS) [25]. When a sufficient magnetic field is applied parallel to the conducting plane, the electron wave function will localize within the plane and the dimensional crossover occurs. As a result, orbital pair breaking weakens and finally vanishes with a further increase of the magnetic field. Therefore, this upward curvature would have to be observed in $H_{c2//}$, which is not observed in our case. Furthermore, the estimated coherence lengths from our data, which are discussed later, show there is no indication of dimensional crossover.

The interesting behavior in this $H_{c2}(T)$ curve is that the two orientations have $H_{c2}$ values that appear to meet at $T = 0$ K. We assume that it is just a coincidence having the close values of $H_{c2}$ for two orientations at the low temperature, where $H_{c2//}$ shows suppression below $T \sim 10$ K by paramagnetic limit, and $H_{c2\perp}$ is limited by orbital effects all the way down to $T = 0$ K without reaching the paramagnetic limit. The temperature dependence of anisotropy ($\gamma$ vs $T$ plot) is shown in the inset of Fig. 5, where anisotropic parameter, $\gamma$, corresponds to $H_{c2//} / H_{c2\perp}$. When determined from pulsed field measurements, $\gamma$ decreases form 2.7 ($T = 18$ K) to 1.1 ($T = 0.7$ K). When determined from PPMS measurements, $\gamma$ shows a maximum value of 3.4 at $T = 21$ K with decreasing temperature. Although a smaller peak value of $\gamma$ is observed near $T_c$ for pulsed field data, the behavior is consistent with data from PPMS measurements determined by the temperature and resistivity accuracy. Our plot of $\gamma$ vs $T$ shows the peak at $T = 0.81 \sim 0.95\ T_c$ which is consistent with theoretical research that expects to have the peak at $T \sim 0.9\ T_c$ in the two-band model [26].

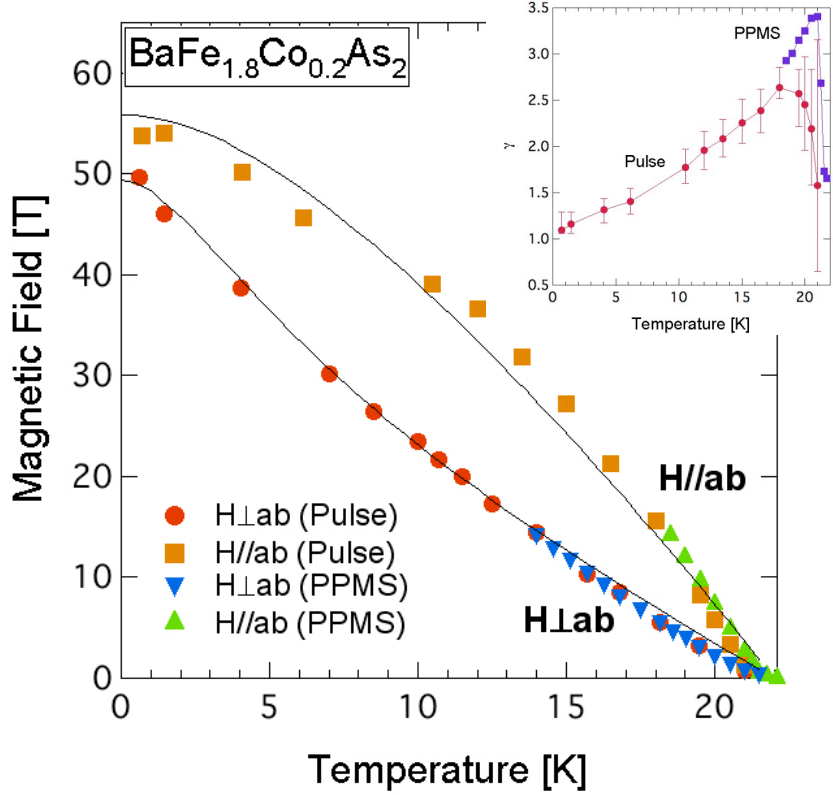

Fig. 5. *H-T* phase diagram in BaFe$_{1.84}$Co$_{0.16}$As$_2$ for field perpendicular to ab-plane and c axis. The solid curves are the 2-band fit (see the text). The inset is the temperature dependence of anisotropy ($\gamma$ vs *T* plot).

Here we obtain the temperature dependence of the anisotropic coherence length to estimate the dimensionality of the superconductivity for in-plane ($\xi_{//}$) and inter-plane ($\xi_\perp$) directions, which is calculated by $\xi_{//} = (\phi_0/2\pi H_{C2\perp})^{1/2}$ and $\xi_\perp = \phi_0/2\pi\xi_{//}H_{C2//}$, respectively. The coherence lengths at $T = 0.7$ K are $\xi_{//} = 25.8$ Å and $\xi_\perp = 24.7$ Å, while both of them are 319 Å at $T = 22$ K. The transverse coherence length ($\xi_\perp$) is significantly longer than the thickness of the conducting sheet d = *c* / 2 = 6.49 Å [4] indicating that the superconductivity is not weakly-coupled two-dimensional (2D) but three-dimensional (3D) in whole temperature range. Thus, there is no decoupling of the superconducting layers. With decreasing temperature, the coherence lengths decreases but $\xi_\perp$ seems to reach a constant value below ~ 10 K, however $\xi_\perp$ might be estimated to be smaller than its true value due to competition with its paramagnetic limit at low temperatures.

Conclusions

We have demonstrated the temperature dependence of the superconducting upper critical field in a BaFe$_{1.84}$Co$_{0.16}$As$_2$ single crystal for the inter- and in-plane directions, which was determined via resistivity. $H_{c2//}(0)$ and $H_{c2\perp}(0)$ are ~ 55 T and ~ 50 T respectively, which is much lower than $H_{c2//}(0)$ extrapolated to zero temperature, of ~ 70 T in the same compound reported earlier [11]. However the phase diagram presented here is consistent with that in Ref. 26. We have shown that for the $H_{c2//}$ curvature, the enhanced paramagnetic limit is dominant at low temperatures, which indicates this is an unconventional superconductor, and that $H_{c2\perp}(0)$ exhibits an upturn near $T = 0$ K, quite possibly due to a multiband feature of this compound. From the fit to the temperature dependence of $H_{c2}$, we showed that this system could be also explained well with two-band model. We have also determined an anisotropy factor having a peak structure at $T \sim 0.9\ T_c$ which is consistent with earlier theoretical predictions of two-band model. Anisotropic coherence lengths for two crystallographic axes prove that this compound is a 3D superconductor.

Acknowledgement

  Research at ORNL sponsored by the Division of Materials Sciences and Engineering, Office of Basic Energy sciences, U.S. Department of Energy.